\documentclass[preprint,5p,times]{elsarticle}

\usepackage{graphics}
\usepackage{amssymb}
\usepackage{lineno}

\usepackage{multirow} %&&&&&&&&&&&&&&&&&&&&&&&&&&&&&&&&&&
\usepackage{hhline}% http://ctan.org/pkg/hhline

\usepackage[utf8]{inputenc}
\usepackage{fancyvrb}

\usepackage{hyperref}

\usepackage{booktabs}

\journal{Fusion Engineering and Design}

\begin{document}

\begin{frontmatter}



\title{Validation of equilibrium tools on the COMPASS tokamak}

 \author[label1]{J.~Urban}
 \author[label2]{L.C.~Appel}
 \author[label3]{J.F.~Artaud} 
 \author[label4]{B.~Faugeras} 
 \author[label1,label5]{J.~Havlicek}
 \author[label1]{M.~Komm}
 \author[label2]{I.~Lupelli} 
 \author[label1,label5]{M.~Peterka} 
 \address[label1]{Institute of Plasma Physics ASCR, Za Slovankou 3, 182 00 Praha 8, Czech Republic}
 \address[label2]{CCFE, Culham Science Centre, Abingdon, Oxfordshire, UK}
 \address[label3]{CEA, IRFM, F-13108 Saint Paul Lez Durance, France}
 \address[label4]{Laboratoire J.A. Dieudonné, UMR 7351, Université de Nice Sophia-Antipolis, Parc Valrose, 06108
Nice Cedex 02, France}
 \address[label5]{Department of Surface and Plasma Science, Faculty of Mathematics and Physics, Charles University in Prague, V Hole\v{s}ovi\v{c}k\'ach 2, 180~00 Praha 8, Czech Republic}

%!TEX root = JUrban_SOFT2014.tex

\begin{abstract}

Various MHD (magnetohydrodynamic) equilibrium tools, some of which being recently developed or considerably updated, are used on the COMPASS tokamak at IPP Prague. MHD equilibrium is a fundamental property of the tokamak plasma, whose knowledge is required for many diagnostics and modelling tools. Proper benchmarking and validation of equilibrium tools is thus key for interpreting and planning tokamak experiments. We present here benchmarks and comparisons to experimental data of the EFIT++ reconstruction code [L.C. Appel et al., EPS 2006, P2.184], the free-boundary equilibrium code FREEBIE  [J.-F. Artaud, S.H. Kim, EPS 2012, P4.023], and a rapid plasma boundary reconstruction code VacTH [B. Faugeras et al., PPCF 2014, accepted]. We demonstrate that FREEBIE can calculate the equilibrium and corresponding poloidal field (PF) coils currents consistently with EFIT++ reconstructions from experimental data. Both EFIT++ and VacTH can reconstruct equilibria generated by FREEBIE from synthetic, optionally noisy diagnostic data. Hence, VacTH is suitable for real-time control. Optimum reconstruction parameters are estimated.

\end{abstract}


\end{frontmatter}

%!TEX root = JUrban_SOFT2014.tex

\section{Introduction} % (fold)
\label{sec:introduction}

We report here on validation and verification of tokamak equilibrium tools used for the COMPASS tokamak \cite{compass2006}. We particularly focus on fundamental global plasma parameters and the shapes of magnetic flux surfaces, which are crucial in diagnostics interpretation and other analyses. 
EFIT++ \cite{efitpp2006} is used for routine equilibrium reconstruction on COMPASS. FREEBIE \cite{freebie2012} is a recent free-boundary equilibrium code; FREEBIE enables predictive equilibrium calculation consistent with the poloidal field (PF) components of the tokamak. In this study, FREEBIE is used in the so-called inverse mode, which predicts PF coils currents from a give plasma boundary and profiles. The third code employed in this study is VacTH \cite{vacthref}, which provides a fast reconstruction of the plasma boundary from magnetic measurements using a toroidal harmonics basis.

In order to verify and validate the aforementioned tools, we analyse EFIT++ and VacTH reconstructions of equilibria constructed with FREEBIE. Synthetic diagnostics (e.g., magnetic probes or flux loops) with optional artificial errors provide inputs for the reconstructions. 

% section introduction (end)

%!TEX root = JUrban_SOFT2014.tex

\section{Verification and validation procedure} % (fold)
\label{sec:procedure}

Reliable MHD equilibrium reconstruction is very important for tokamak exploitation. Numerous diagnostics and subsequent analyses require as inputs equilibrium properties such as flux surface geometry, magnetic field, stored energy, internal inductance etc. We have set up a set of benchmarking tasks, which verify and validate equilibrium tools that are currently employed on COMPASS. The procedure is fundamentally following:

\begin{enumerate}
	\item Equilibrium reconstruction of selected experimental cases using EFIT++.
	\item Recalculate the equilibria using FREEBIE in inverse mode.
	\item Optionally alter the equilibria in FREEBIE using e.g. experimental pressure profiles.
	\item Reconstruct FREEBIE equilibria using EFIT++ and VacTH with various parameters and artificial input noise.
\end{enumerate}

The first step employs a routine EFIT++ set-up for COMPASS with heuristically tuned parameters. In addition to the total plasma current $I_\mathrm{p}$ and the currents in individual PF circuits, 16 partial Rogowski coils and 4 flux loops are employed in this reconstruction and $p'$ and $FF'$ are assumed to be linear functions of the poloidal flux $\psi$.

In the second step, FREEBIE inputs $I_\mathrm{p}$, $p'\left( {\bar \psi } \right)$ and $FF'\left( {\bar \psi } \right)$ profiles, the plasma boundary coordinates and an initial guess for the PF coils currents. Here, $p$ is the plasma pressure, $F = RB_\phi$ and $\bar\psi$ is the normalized poloidal magnetic flux ($\bar\psi = 0$ on the magnetic axis and $\bar\psi = 1$ on the plasma boundary). $p'$ comes either from the EFIT++ reconstruction or from Thomson scattering pressure profile $p_\mathrm{TS} = 1.3 n_\mathrm{e} p_\mathrm{e}$. FREEBIE then seeks a solution to the Grad-Shafranov equation, including the PF coils currents, which minimizes the given plasma shape constraint. (This regime is called the inverse mode.)

It should be noted here that to set up a free-boundary equilibrium code, a rather complete machine description is necessary (in particular, the PF coils geometry and circuits, limiter, vessel and other passive PF elements and magnetic diagnostics configuration). We adopted the description that was already available for EFIT++ and transformed it to ITM CPO's (Integrated Tokamak Modelling Consistent Physical Objects) structures, which are subsequently either used directly in FREEBIE or converted to VacTH specific input format.

FREEBIE can naturally output arbitrary synthetic diagnostics. We use here additional 24 poloidally and 24 radially oriented partial Rogowski coils (which are actually mounted on COMPASS) and an artificial set of 16 flux loops located at the same positions as the basic magnetic probes. Hereafter, the number of magnetic probes and flux loops are denoted $n_\mathrm{mp}$ and $n_\mathrm{fl}$. $n_\mathrm{mp}=16$, $n_\mathrm{fl}=4$ refers the basic set of magnetic measurements, $n_\mathrm{mp}=64$ refers to a set of all presently mounted partial Rogowski coils on COMPASS and $n_\mathrm{fl}=16$ implies artificial flux loops positioned at the same locations as the basic magnetic probes.

In the optional third step, an artificial random noise is added to the calculated values of $I_\mathrm{p}$, magnetic probes and flux loops. In particular, for a given noise level $\epsilon$, $\tilde X = \left( {1 + U\left( { - \epsilon, \epsilon} \right)^\mathrm{T}} \right)X$, where $X$ is a row vector of the synthetic diagnostics data and $U\left( { - \epsilon, \epsilon} \right)$ is a random vector of the same shape as $X$ with a uniform distribution on $\left( { - \epsilon, \epsilon} \right)$.

The final fourth step consists of reconstructing the equilibria form synthetic FREEBIE data using EFIT++ and VacTH. The reconstructions are then compared to the original equilibrium, focusing on global parameters and geometry. Scans are performed over noise levels ($\epsilon$) and selected code parameters: $p'$ and $FF'$ polynomial degrees in EFIT++ ($n_{p'}$, $n_{FF'}$) and the number of harmonics ($n_P$, $n_Q$) in VacTH. The following quantities are used for the comparison.
\begin{table}[!h]
    \begin{tabular}{ll}
    $R_{\mathrm ax}$, $Z_{\mathrm ax}$ & $R,Z$ coordinates of the magnetic axis \\
    $R_{\mathrm in}$, $R_{\mathrm out}$ & inner/outer $R$ coordinate of LCFS at $Z=Z_{\mathrm ax}$ \\
    $Z_{\mathrm min}$, $Z_{\mathrm max}$ & minimum/maximum $Z$ coordinate of LCFS \\
    $I_{\rm{p}}$ & plasma current \\
    $\kappa  = \frac{\left( {{Z_{{\rm{max}}}} - {Z_{{\rm{min}}}}} \right) } { \left( {{R_{{\rm{out}}}} - {R_{{\rm{in}}}}} \right) }$          & elongation   \\
    ${l_{\rm{i}}} = {{\bar B_{\rm{p}}^2}} / {{B_{\rm{a}}^2}}$          & normalized internal inductance   \\
    ${\beta _{\rm{p}}} = {{2{\mu _0}\bar p}} / {{B_{\rm{a}}^2}}$          & poloidal beta   \\
    $ W = \int_0^V {\frac{3}{2}p{\rm{d}}V'}$ & stored plasma energy \\
    $ q_0$, $q_{95}$ & safety factor at $\bar \psi = 0,\ 0.95$ \\
    \end{tabular}
\end{table}

Here, $\bar x = \int_0^V {x/V{\rm{d}}V'} $ is a volume average, $V$ is the total plasma volume, $B_{\mathrm a} = \mu _0 I_\mathrm{p} / l_{\mathrm a}$, $l_{\mathrm a}$ is the poloidal LCFS (last closed flux surface) perimeter.
We also define absolute differences $\Delta x = {{x_0} - x} $ and relative differences $\delta x = \left| {{x_0} - x} \right|/\left| {{x_0}} \right|$, where $x$ is an arbitrary recontructed quantity and $x_0$ its target value.

VacTH does not provide a full equilibrium but the plasma shape only as the target of VacTH is to provide such reconstructions in real time for a feedback control. The inputs (besides the machine description and code parameters) of VacTH are PF coil currents, $I_{\rm{p}}$, magnetic probes and flux loops measurements and the initial axis and X-point coordinates (the coordinates can be fixed as code parameters).

% section procedure (end)

%!TEX root = JUrban_SOFT2014.tex

\section{Results} % (fold)
\label{sec:results}

\begin{figure*}[!htb]
\centering   %\begin{center}
\hfill{}
\includegraphics[width=18cm]{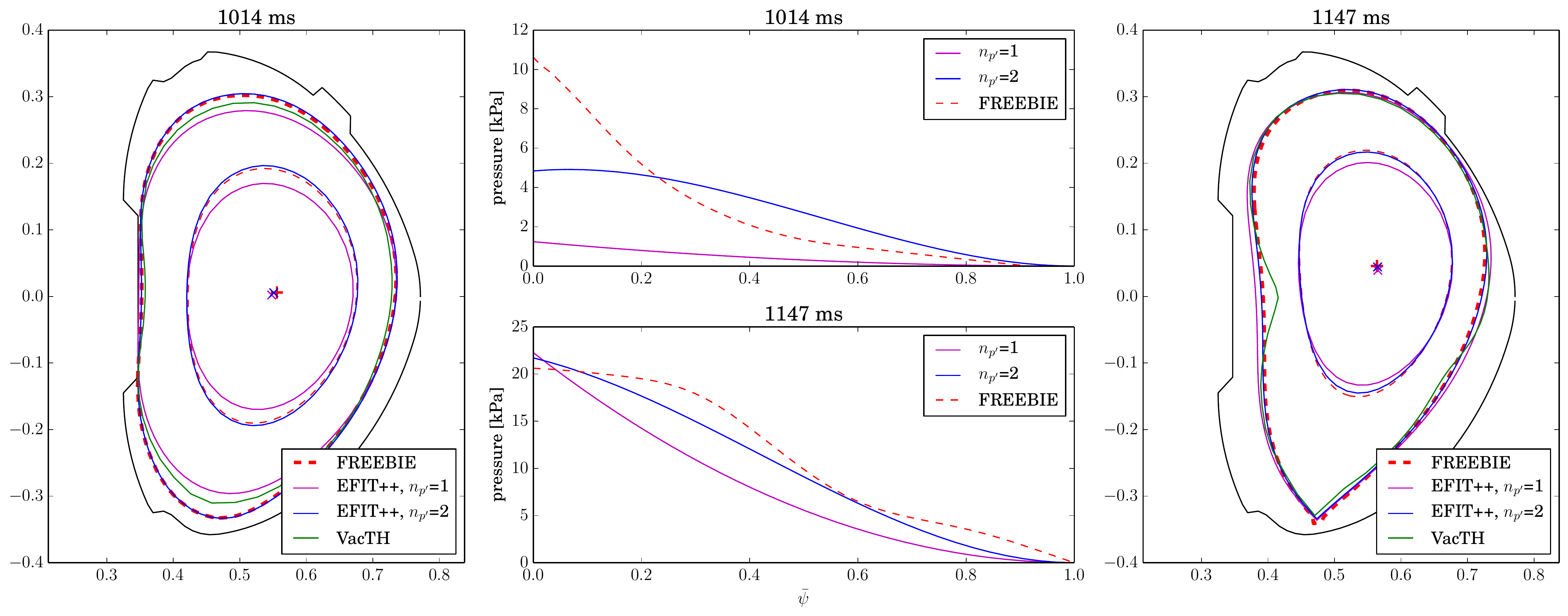}
\hfill{}
%\end{center}
\caption{EFIT++ reconstructed pressure profiles and contours of $\bar\psi=\left(0.5,1\right)$ and VacTH LCFS from FREEBIE data, shot 6962 with Thomson scattering pressure profiles. EFIT++ parameters: $n_\mathrm{mp} = 16$, $n_\mathrm{fl} = 4$, $n_{FF'} = 1$. VacTH parameters: $n_\mathrm{mp} = 8$, $n_\mathrm{fl} = 16$, $n_P = n_Q = 4$. 3~\% random input data noise is used in the case of VacTH and EFIT++ with $n_{p'} = 2$, zero noise otherwise.}
\label{fig:ex6962}
\end{figure*}

We have selected five time slices from COMPAS shots 4275 and 6962 (i.e. 10 cases in total) for the analysis. These cases include circular, elongated and diverted plasmas with different currents. 

\subsection{Example cases}

Examples of EFIT++ and VacTH reconstructions are shown in Fig. \ref{fig:ex6962}. The results are quite typical. EFIT++ with a linear $p'$ yields an enhanced LCFS error, in particular in the first, elliptical plasma case, even with zero input data noise. On the other hand, quadratic $p'$ reconstructs the plasma shape correctly even with a noisy input. A similar observation applies to the pressure profiles, except that for the elliptical case, the pressure is not well reconstructed for either $n_{p'}$. 

VacTH reconstructs the LCFS reasonably, even with noisy inputs. Although a bending artefact emerge on the inboard side. Similar artefacts can be observed in other VacTH results as well. This is probably a result of the specific COMPASS configuration as such a behaviour was not observed in the case of WEST \cite{vacthref}. $n_P = n_Q = 4$ is used in this case as these values are minimum for reasonable VacTH results, while higher values are too sensitive to the input noise.

\begin{figure*}[!htb]
\centering   %\begin{center}
\hfill{}
\includegraphics[width=18cm]{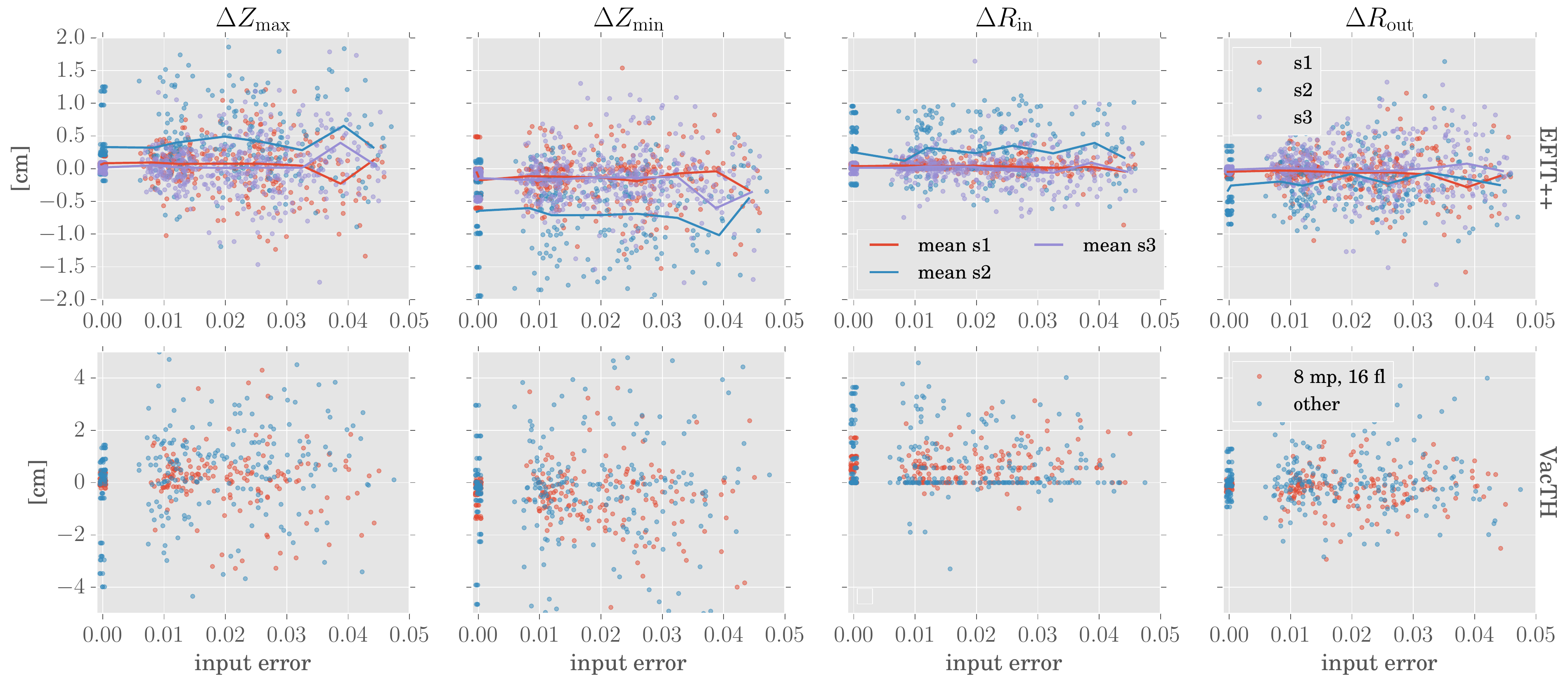}
\hfill{}
%\end{center}
\caption{Absolute errors in LCFS extents for convergent cases. EFIT++ results in the first row, VacTH in the second row. S1 denotes linear $p'$ and $FF'$ in EFIT++ as well as FREEBIE, s2 denotes TS pressure profiles in FREEBIE and $n_{p'} = n_{FF'} = 1$ in EFIT++, s3 denotes TS pressure profiles in FREEBIE and $n_{p'} = 2$ in EFIT++. Full lines show the means. VacTH parameters are $n_P=n_Q=4$, ``8 mp, 16 fl'' denotes $n_\mathrm{mp}=8$, $n_\mathrm{fl}=16$. Input error is calculated as an average of $I_{\rm{p}}$, magnetic probes and flux loops values. Zero input error data are scattered for a better visibility.}
\label{fig:RZstats}
\end{figure*}

\begin{figure*}[!htb]
\centering   %\begin{center}
\hfill{}
\includegraphics[width=18cm]{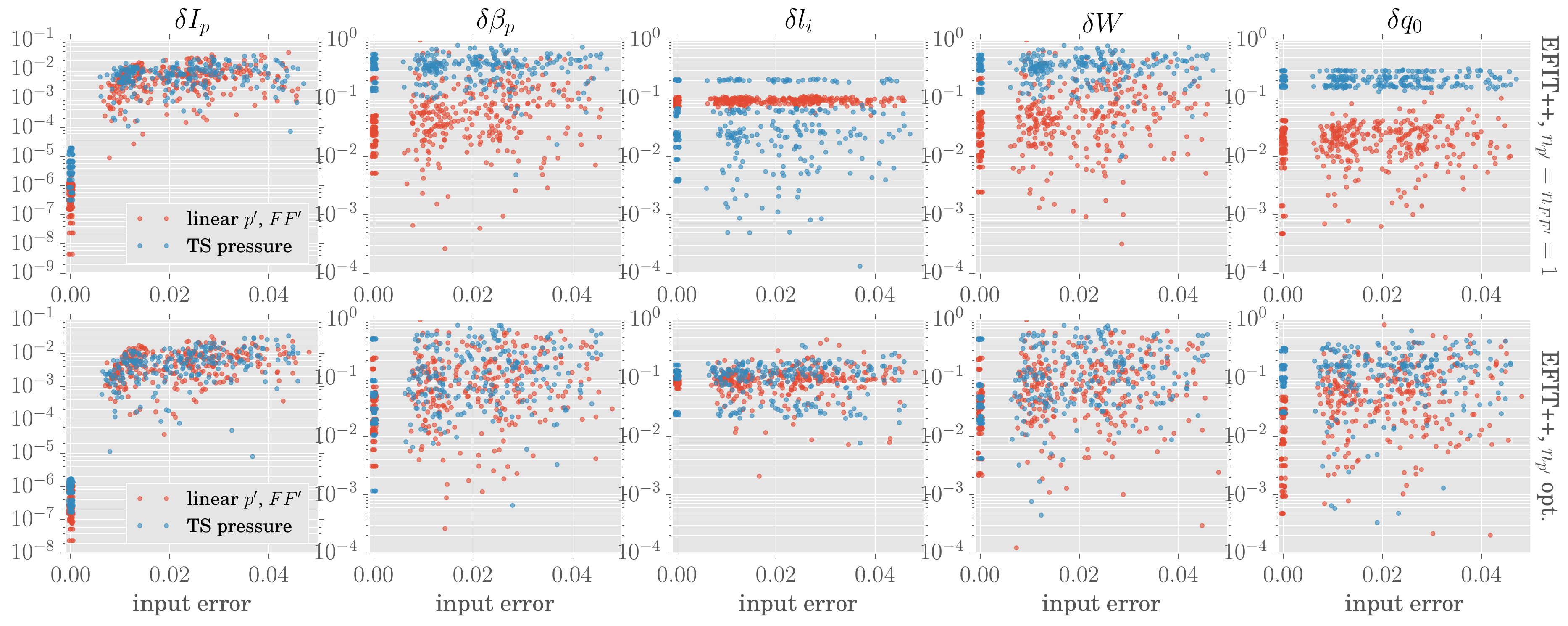}
\hfill{}
%\end{center}
\caption{Internal plasma parameters relative errors for EFIT++ reconstructions using $n_{p'}=n_{FF'}=1$ in the top row and more optimized $n_{p'}$ in the bottom row. Zero input error data are scattered for a better visibility.}
\label{fig:kinetic_stats}
\end{figure*}

\subsection{Statistical analysis} % (fold)
\label{sub:statistical_analysis}

In order to get a global overview of EFIT++ and VacTH reconstruction properties on COMPASS, we perform a scan over major code parameters and signal noise levels. In particular, $n_{p',FF'} = 1,2 $, $(n_\mathrm{mp}, n_\mathrm{fl}) = (16, 4), (64, 4), (8, 16)$, $\epsilon = 0, 0.02, 0.04, 0.06$, $n_{P,Q} = 4, 5, 6$. The same cases as above (time slices of shots 4275 and 6962) are used as target equilibria. For shot 6962, equilibria with TS pressure profiles and with linear $p'$ and $FF'$. This means there is 15 different target equilibria in total.

Absolute errors of the reconstructed LCFS extents for convergent cases from the scan are shown in Fig. \ref{fig:RZstats}. We can observer that the LCFS reconstructed with EFIT++ for target linear $p'$ and $FF'$ profiles (selection 1) are within 1~cm errors. There are, however, cases with up to 3 cm errors in $Z_{\rm{min}}$ for the more realistic TS pressure profiles (selection 2) if $n_{p'}=1$ is used. This error can be reduced by using $n_{p'}=2$. Input errors do not pose major difficulties for EFIT++. 

VacTH is performing reasonably well for its most favourable diagnostic set of 8 magnetic probes and 16 flux loops and $n_P = n_Q = 4$. With a higher number of harmonics or with less flux loops, VacTH becomes unreliable and yields significant errors. Unfortunately, only 4 flux loops are currently available on COMPASS. In fact, it is easier for VacTH to fit magnetic probes than flux loops
An additional optimization of the fitting weights or algorithm is probably needed. The current behaviour might be quite anti-intuitive as VacTH performs significantly worse with 16 flux loops and 16 or 64 magnetic probes in comparison to 16 flux loops and only 8 magnetic probes. 

EFIT++ internal plasma parameters reconstruction results are shown in Fig. \ref{fig:kinetic_stats}. It shows that purely magnetic reconstruction with $n_{p'}=n_{FF'}=1$ introduces (except for $I_{\rm{p}}$) a systematic error for realistic pressure profiles, i.e. for plasmas that do not have the same profile parametrization. 
It is known that magnetic reconstruction with EFIT is difficult for small circular plasmas (without additional constraints, particularly the stored energy) \cite{efit1985}.
This suggests that using $n_{p'}=1$ for circular plasmas and $n_{p'}=2$ for elongated and diverted plasmas might lead to better results. This is demonstrated in the bottom row of Fig. \ref{fig:kinetic_stats}.
Reconstructions with such optimized parameters do not suffer from the systematic error; however, they generally increase the error bars for target equilibria with linear $p'$ and $FF'$, especially for $q_0$. It is also notable that $\delta l_{\rm{i}} \cong 0.1$ for all $n_{p'}=n_{FF'}=1$ reconstructions.

\begin{figure}
\centering   %\begin{center}
\hfill{}
\includegraphics[width=8cm]{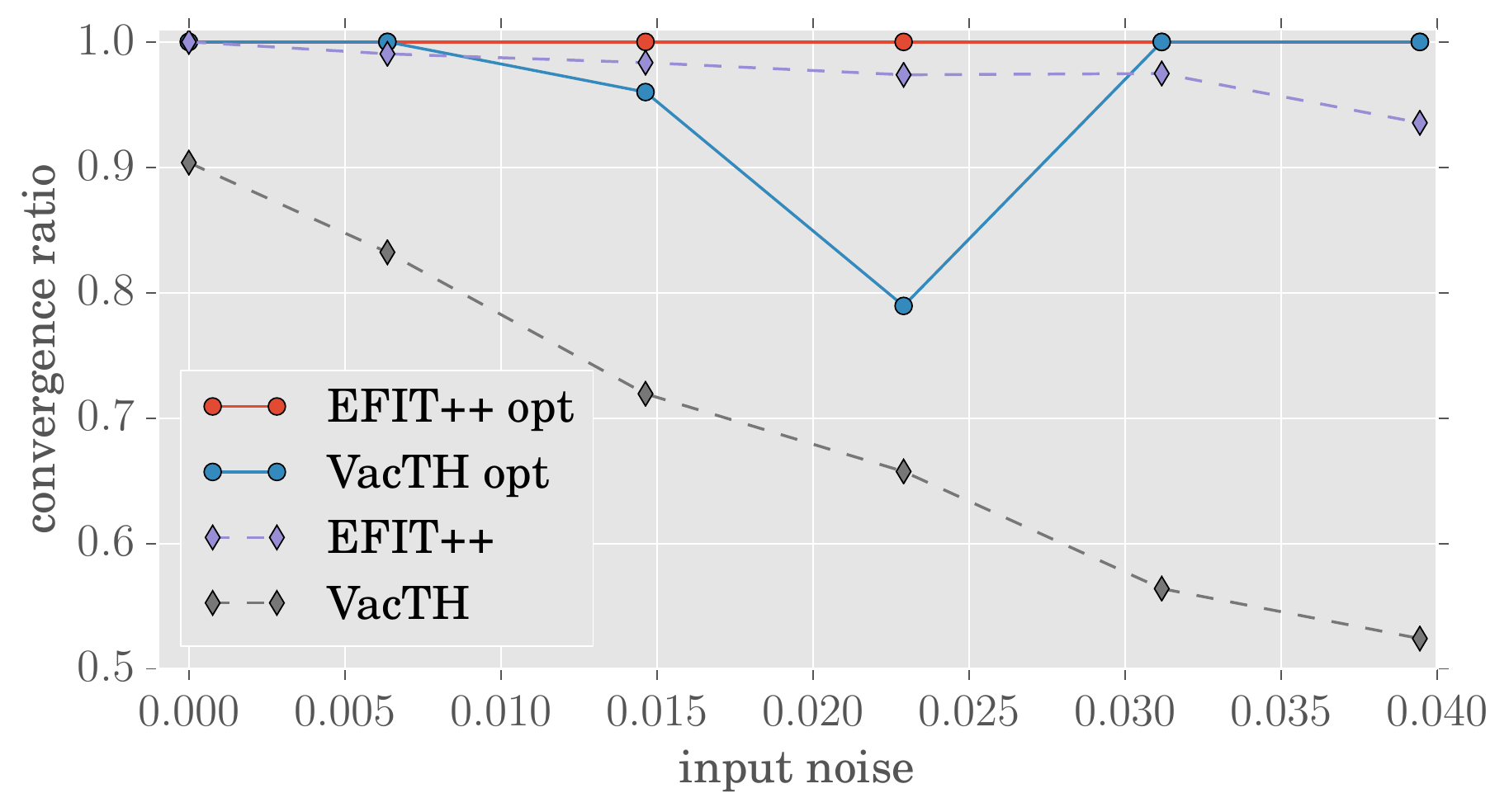}
\hfill{}
%\end{center}
\caption{Ratio of converged cases to all cases, shot 6962, TS profiles. Opt refers to optimized code parameters.}
\label{fig:convergence_ratio}
\end{figure}

Another important property is the converged cases ratio, shown in Fig. \ref{fig:convergence_ratio}. EFIT++ converges in almost all cases with any of the tested configurations and in 100~\% cases in the optimized configuration (i.e. with $n_{p'}=2$ for diverted plasmas). $n_P = n_Q = 4$ must be used in VacTH unless the number of non-converged cases is too large. 
Quite interestingly, optimized VacTH convergence rate drops significantly around 2~\% input noise.

% subsection statistical_analysis (end)

% section results (end)

%!TEX root = JUrban_SOFT2014.tex

\section{Conclusions} % (fold)
\label{sec:conclusions}

Two new codes---FREEBIE and VacTH---have been successfully set up on COMPASS, which enabled to perform an extensive cross-benchmarking and validation of free-boundary equilibrium tools. We show that FREEBIE can predict equilibria that are consistent with EFIT++ reconstructions from experimental data. FREEBIE model equilibria, either with linear $p'$ and $FF'$ profiles or with pressure profiles from Thomson scattering diagnostic, then served to assess the credibility of EFIT++ reconstructions. 

We show that magnetic reconstruction EFIT++ with linear $p'$ and $FF'$ features a relatively good accuracy of 1 -- 2 cm in the plasma shape reconstruction but introduces systematic errors both in the shape and in internal plasma parameters, such as $W$, $l_{\mathrm i}$, $\beta_{\mathrm p}$ or $q_0$. The reconstruction properties can be significantly improved by using quadratic $p'$ for elongated and divertor plasmas, which removes the systematic error and also improves the LCFS reconstruction. EFIT++ converges in 100~\% cases in this regime.

Optimum parameters for VacTH have been estimated. In particular, the optimum number of harmonics is 4 otherwise VacTH fails to converge in many cases, even without any input error. 16 flux loops and only 8 magnetic must be used as VacTH input. With less flux loops or more magnetic probes the code performs significantly worse. We conclude that VacTH is a promising tool pertinent for a real-time feedback control of the plasma shape.

% section conclusions (end)

\section*{Acknowledgement}
The work of J. Urban was supported by Czech Science Foundation grant 13-38121P.
The work at IPP ASCR was supported by MSMT \#LM2011021. 

\bibliographystyle{elsarticle-num}
\bibliography{biblio}{}

\end{document}